\documentclass[twocolumn,showpacs,preprintnumbers,amsmath,amssymb]{revtex4}

\usepackage{graphicx}
\usepackage{dcolumn}
\usepackage{bm}

\begin{document}

\title{\large Magnetoresistance of
Junctions made of Underdoped YBCO Separated by a Ga-doped YBCO
Barrier}

\author{L. Shkedy, G. Koren and E. Polturak}
\affiliation{Physics Department, Technion - Israel Institute of Technology Haifa, 32000, ISRAEL}

 \email{lior_shk@physics.technion.ac.il}

\date{\today}

\begin{abstract}
We report magnetoresistance measurements of ramp type
superconductor-normal-superconductor (SNS) junctions. The
junctions consist of underdoped $YBa_{2}Cu_{3}O_y$ (YBCO)
electrodes separated by a barrier of $YBa_{2}Cu_{2.6}Ga_{0.4}O_y$.
We observe a large positive magnetoresistance, linear in the
field. We suggest that this unusual magnetoresistance originates
in the field dependence of the proximity effect. Our results
indicate that in underdoped YBCO/N/YBCO SNS structures, the
proximity effect does not exhibit the anomalously long range found
in optimally doped YBCO structures. From our data we obtain the
diffusion coefficient and relaxation time of quasiparticles in
underdoped YBCO.\\

\pacs{74.45.+c , 74.25.Ha}

\end{abstract}

\maketitle{\large   INTRODUCTION}

In the usual description of the proximity effect, when a
superconductor (S) is brought into contact with a normal conductor
(N), the order parameter (OP) in the superconductor is depressed
near the interface and superconductivity is induced in N. The pair
amplitude induced in N decays on a length scale $K^{-1}$ from the
interface, called the decay
length\cite{RMPdeGennes,Parks,Tinkham}. In SNS junctions in which
S is an optimally doped HTSC and N belongs to the same material
family, but is doped to be non superconducting, the decay of the
pair amplitude in N typically takes place over a rather long
distance of tens of nm,\cite{Kleinsasser,Sharoni2004}. In
contrast, if both S and N are underdoped cuprates, the pair
amplitude in N seems to decay over a much shorter distance, on the
order of a few nm. We have observed this effect in  underdoped
YBCO based junctions\cite{APLofer,nesher2} having a barrier made
of $YBa_{2}Cu_{2.55}Fe_{0.5}O_y$ or a
$YBa_{2}Cu_{2.6}Ga_{0.4}O_y$. In SNS junctions having a barrier
much thicker than the decay length, Cooper pairs cannot tunnel
through and the junctions exhibit a finite resistance at all
temperatures. Roughly speaking, superconductivity in N is induced
near the two SN interfaces, while a section of length $\ell$ in
the middle of the barrier remains normal. This is the type of
junction studied in the present work.

We are not aware of previous investigations of the proximity effect
in HTSC under a magnetic field. When a magnetic field is
applied, superconductivity is reduced and penetrates less into the
normal conductor. As a result, the proximity effect is field
dependent\cite{Parks}. If the superconductivity in the barrier is
weakened, the length of the normal section in the junction
should increase, and with it the junction's finite resistance. As
a result, a positive magnetoresistance (MR, defined as $MR\equiv
R(H)-R(0)$) should be observed. We indeed observed such MR, linear
in the field. An attempt to explain this unusual field dependence
is the subject of this paper.

Besides the field dependence of the proximity effect,
there are several additional
mechanisms which could contribute to the MR. These include flux
flow in the superconducting electrodes
\cite{BardeenStephan,Tinkham}, normal MR of the barrier material
which is caused by bending of electron trajectories \cite{Ziman},
field dependent hopping in the barrier \cite{Lifshitz,Shklovskii},
and resonant tunneling between the electrodes across the barrier
\cite{Abrikosov}. In the following we show that the contribution
of all these processes to the observed MR is insignificant and we
attribute it primarily to a
field dependent proximity effect in the barrier.\\

\maketitle{\large   EXPERIMENTAL}

The junctions used in the present study are thin film based ramp
junctions of the type that was previously used in our work
\cite{APLofer}. The junctions consist of two underdoped
superconducting YBCO electrodes separated by a thin layer of
Ga-doped YBCO barrier. Ga has no magnetic properties. The
transport current flows in the a-b plane through the Ga-doped YBCO
layer. The multi-step process of junction preparation by laser
ablation was described previously\cite{APLofer}. Briefly, we first
deposit a 100 nm thick c-axis oriented epitaxial YBCO layer  onto
a (100) $SrTiO_3$ (STO) substrate. This base electrode is then
capped by a thick insulating layer of STO. Patterning is done by
Ar ion milling to create shallow  angle ramps along a main
crystallographic direction in the a-b plane. In a second
deposition step, the barrier layer, the YBCO cover electrode and
the Au electrical contacts are deposited, and then patterned to
form the final junctions layout. This produces several junctions
with 5$\mu$m width on the wafer. Four terminal resistance
measurements  of the junctions were carried out as a function of
temperature and magnetic field of up to 8 Tesla. The field was
perpendicular to the transport current, which in our geometry
flows in the a-b plane of the films.\\

\begin{figure}
\includegraphics[height=7cm,width=9cm]{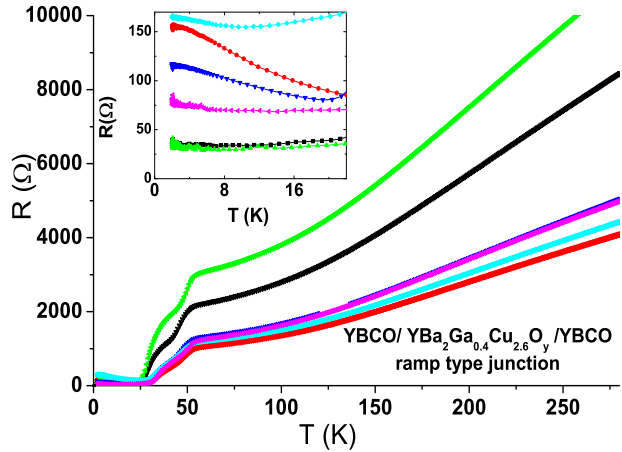}
\caption{\label{fig:epsart} (Color online) Resistance versus
temperature of six junctions with 21nm thick Ga-doped YBCO
barrier. In the normal state, the different resistances  of the
junctions are due to different lengths of the YBCO leads. The
inset shows the low temperature resistance of the junctions where
both electrodes are superconducting. }
\end{figure}

\begin{figure}
\includegraphics[height=7cm,width=9cm]{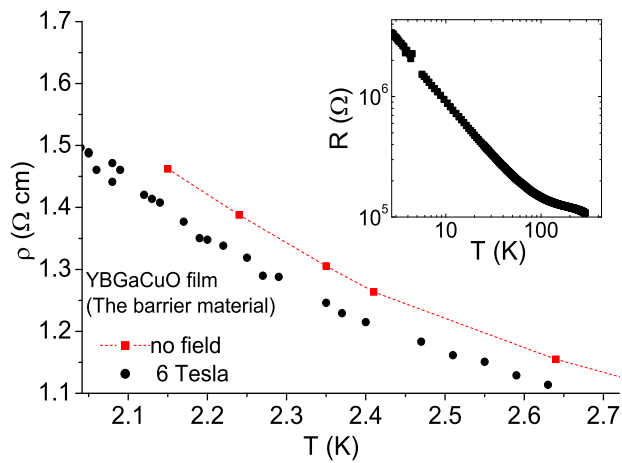}
\caption{\label{fig:epsart}  (Color online) Resistivity versus
temperature of ~100nm thick film of the Ga-doped YBCO material.
Square symbols are measured at zero field and the circles are
measured with 6 Tesla field applied perpendicular to the film.
Note that the MR of the film is negative, in contrast to the
positive MR of our junctions.}
\end{figure}

\vspace{1cm}

\maketitle{\large RESULTS AND DISCUSSION}

Resistance versus temperature (RT) measurements of six junctions
on the wafer are shown in Fig. 1. In the normal region, the
difference in the resistance of the junctions is due to the
different lengths of the YBCO leads. One observes two distinct
superconducting transitions with $T_c$ onset of 35K and 53K which
are attributed to each one of the electrodes. In the oxygen
annealing process of underdoped YBCO, the oxygen concentration is
kept low and the duration of the annealing is relatively short.
Consequently, the base electrode which is covered by a thick layer
of STO, absorbs less oxygen and its transition temperature is
lower.  Below about 30K, both electrodes are superconducting and
the inset of Fig.1 shows the low temperature resistance of the
junctions, which is due to the barrier. Qualitatively similar
behaviour was observed in edge junctions made of underdoped YBCO
separated by a $YBa_{2}Cu_{2.55}Fe_{0.5}O_y$
barrier\cite{APLofer,nesher2}. The scatter of the values between
different junctions is typical of our junction preparation
process, and is probably due to nonuniformities in the local Ga
concentration, and to variations in the transparency of the
interfaces, most probably resulting from damage created by the ion
milling of the ramp. The transparency of our junctions can be
estimated from measurements of the critical current described
below, which indicate that the transparency is low. The
temperature dependence of the junction's resistance is typically
weaker than that of the parent material in the form of a film,
shown in Fig. 2. At low temperatures, the absolute resistivity of
the junctions is also much smaller than that of a
$YBa_{2}Cu_{2.6}Ga_{0.4}O_y$ film. One possible interpretation is
that the thickness of the barrier (21 nm in this work) is in the
range where the material is mesoscopic. Under these conditions,
the temperature dependence of the resistance is expected to be
much weaker than that of a macroscopic film\cite{Glazman}. The
differences of the absolute resistivities between different
junctions may perhaps result from different interface
transparency, which also affect the conductance of the
device in the mesoscopic regime\cite{Glazman}.\\

Our main experimental result is shown in Fig. 3, where the
measured magnetoresistance MR at low T is plotted as a function of
magnetic field normal to the wafer. All junctions showed similar
behavior. Detailed measurements were done on three of the six
junctions on the wafer. One can see that all three junctions show
a large positive MR which is linear in the applied field. The MR
typically reaches a value of $\sim 20\Omega$ at 8 Tesla, which is
larger than the resistance at H=0 by tens of percent.\\

\begin{figure}
\includegraphics[height=7cm,width=9cm]{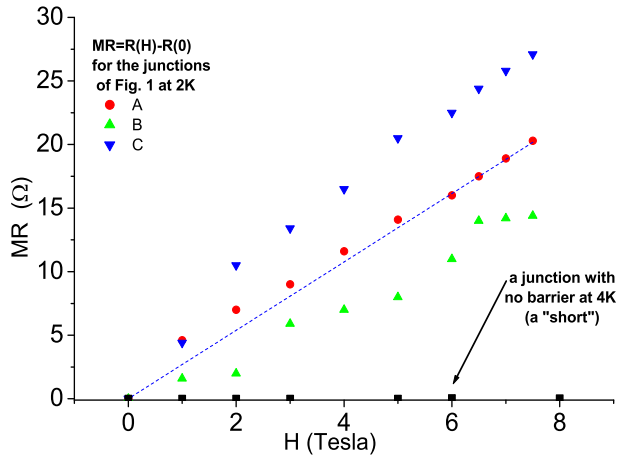}
\caption{\label{fig:epsart} (Color online) Magnetoresistance
versus field of three of the junctions of Fig. 1 at 2K, and of a
"short" junction at 4K. The "short" resistance is of about
0.4$\Omega$ at 8 Tesla which is almost two orders of magnitude
smaller than the corresponding MR of the other junctions with a
barrier. The dashed line is a guide to the eye.}
\end{figure}

We consider possible sources for the MR in our junctions. MR
originating in the two YBCO electrodes below the superconducting
transition temperature can result for instance from flux motion.
This contribution would be linear in the field. In order to
estimate the size of this contribution, we performed low
temperature MR measurements on bare YBCO microbridges. At
temperatures close to $T_c$, flux flow was indeed observed (see
Appendix ). However, at low temperatures where the junctions of
Fig. 3 were measured, no measurable MR  was observed in the thin
film YBCO microbridges. Therefore, flux flow in the YBCO
electrodes does not contribute to the MR. We also measured the MR
of junctions prepared in the same way, but without the barrier
layer. These junctions are referred to as "shorts".  As shown in
Fig. 3, under similar bias currents and fields, the "shorts" did
exhibit a small MR of about $0.4\Omega$ at 8 Tesla. The "shorts"
show a finite MR since the interface between the two YBCO
electrodes is always imperfect. The MR of the "shorts" is smaller
than the MR in the junctions by almost two orders of magnitude.
The interface resistance cannot be directly measured. What can be
measured is the critical current density.  Typically, the critical
current density at low temperature of a 60K YBCO "short" is  one
order of magnitude smaller than that of a film. This implies that
the transparency of our junctions is low. To summarize this
section, the above series of control experiments show that MR in
the electrodes is \textit{not} the source of the large MR observed
in our junctions.\\

A second potential source for the observed MR in Fig. 3 could be
the barrier material itself. We therefore measured the MR of the
Ga-doped YBCO. Specifically, we measured the resistance versus
temperature of microbridges patterned in a thin film of this
material annealled under the same conditions as the junctions in
Fig. 3. Fig. 2 shows the resistivity of these bridges with and
without magnetic field. The barrier material exhibits a clear
\textit{negative} MR  of $\sim$5\% at 2K. The sign of this MR is
opposite to that of the junctions which show a large
\textit{positive} MR. At low temperatures, where the MR of the
Ga-doped films is largest, the MR contributed by the barrier in
the junctions would be at most -8$\Omega$ (5\% of $160\Omega$, as
seen in the inset of Fig. 1). However, since the sign of the MR of
the barrier material itself is negative, the net (positive) MR of
the junctions should be even larger than shown in Fig. 3.
Consequently, the properties of the barrier material on its own
cannot explain the observed MR of the junctions.\\

The above mentioned control experiments clearly show that the MR
of our junctions does not originate from the superconducting
electrodes nor from the normal properties of the barrier material.
The net MR which we see has a magnitude characteristic of the
transition of part of the barrier from a superconducting to a
normal state. We therefore examine whether the MR could originate
from the depression of superconductivity near the SN interface of
the junction.\\

Before going into a more detailed analysis, we note that our
barrier is a mesoscopic section of a Mott insulator (MI), with the
conductance of the material in bulk form showing variable range
hopping \cite{KorenMRO}. Its low temperature resistivity,
$~0.8\Omega cm$, is about 3 orders of magnitude larger than the
maximum resistivity of metals (Mott-Ioffe-Regel limit
\cite{Ioffe}). Strictly speaking, our junctions are S/MI/S
junctions. So, the application of the usual theoretical
description of the proximity effect to our junctions is not a
priori justified, since both the de-Gennes and Usadel equations
are valid only for dirty metals. However, it is an experimental
fact that when an MI with resistivity $\rho \leq 1\,\Omega cm$ is
in good electrical contact with a superconductor it behaves
similarly to a metal \cite{KorenMRO,Hashimoto,Stozel,Frydman}. The
question which particular model to use is therefore a matter of
choice. In the limit of small induced pair amplitude in N, which
applies to our low transparency junctions, the de-Gennes and
Usadel approach give the same result. Since the de-Gennes approach
was traditionally employed in all previous and current work on
HTSC proximity structures\cite{Kleinsasser,Polturak,Bozovic}, we
prefer to follow this route. In any case, the analysis presented
below is nevertheless useful, in terms of assigning values to
physical quantities such as the decay length which can then be
inter-compared between different experiments. \\

We first discuss the MR on the S side of the SN interface. In this
region, the order parameter is reduced, superconductivity is
depressed, pinning is weakened and flux flow could occur despite
the low temperature. We now estimate the upper limit on the
contribution of this effect to the MR. The low temperature normal
state resistivity of YBCO, extrapolated from the linear part of
the RT plot above the transition, is about $10^{-4}\Omega cm$. An
upper limit on the volume near the interface in which
superconductivity is weakened is $10\xi \times A \sim
200\rm\AA$$\times 0.5\mu m^{2}$, where A is the junction cross
section \cite{Lubimova}.  The normal state resistance of this
region is very low, less than $0.1\Omega$. Since the flux flow
resistance is a fraction of the normal state resistance, it
follows that the MR in the S side close to the interface is
negligible.\\

Turning now to the N side of the interface, the resistivity of the
barrier material is quite high, $~0.8\Omega cm$ at 2K. A rough
estimate done assuming Ohm's law in the barrier indicates that a 1
nm thick slice of the barrier has a resistance of $R\sim
16\Omega$. This value is similar to the total MR seen in Fig. 3.
In the following, we propose that the observed MR is caused by
changes in the effective penetration of superconductivity into the
barrier. In other words, when a magnetic field is applied, the
magnitude of the pair amplitude induced in the barrier is
decreased and $\ell$, the effective length of the barrier which
remains normal, increases thus
increasing the resistance of the junction.\\

The magnetic fields used in the present study are small compared
with $H_{c2}$ of the 60K YBCO phase which is ~50T \cite{Ando}.
Thus changes in the minigap $\Delta$ due to the applied field are
also small but not negligible. The value of $\Delta$ on the S side
near the interface is proportional to $T_{c}$ which itself depends
on the magnetic field due to pair breaking according to
\cite{Tinkham,MakiParks}.

\begin{equation}\label{1}
ln\left(\frac{T_{c}}{T_{c0}}\right)=\Psi\left(\frac{1}{2}\right)-\Psi\left(\frac{1}{2}+\frac{\alpha}{2\pi
kT_{c}}\right)
\end{equation}

\noindent where $T_{c}$ is the critical temperature under applied
field and $T_{c0}$ is the critical temperature at zero field.
$\Psi$ is the di-gamma function defined as
$\Psi(x)=\Gamma'(x)/\Gamma(x)$, and $\alpha$ is the pair breaking
parameter. For a thin film under perpendicular applied field
$\alpha=D_{S}eH/c$, where $D_{S}$ is the diffusion coefficient in
the superconductor. Since the highest magnetic field we used is
small compared with $H_{c2}$, pair breaking is small and $(\alpha
/2\pi k_{B}T_{c})$ is a small parameter. In this limit, Eq. (1)
reduces to \cite{Tinkham,MakiParks}

\begin{equation}\label{2}
k_{B}(T_{c0}-T_{c})=\frac{\pi\alpha}{4}
\end{equation}

\noindent From our RT measurements under different fields we find
the values of $T_{c0}$ and $T_c$ (65K at 0T and 55K at 7T,
respectively). We can thus calculate the value of $\alpha$ which
is $\sim$1 meV at 7 Tesla. Therefore, $\alpha /2\pi
k_{B}T_{c}\simeq 1/35$, and this justifies the use of Eq. (2). By
assuming a linear scaling between $\Delta$ and $T_c$
($2\Delta=\beta k_B T_c$, with $\beta$ being a constant of about
5), we estimate that under a field of 7T the magnitude of $\Delta$
decreases by about 15\%. The suppression of $\Delta$ can be
therefore written as

\begin{equation}\label{3}
\delta_S\equiv\Delta_S(0)-\Delta_S(H)=\frac{\pi\alpha\beta}{8}=\frac{\pi\beta}{8}\frac{D_{S}eH}{c}
\end{equation}

\noindent where $\delta_S$ is small compared with $\Delta_S(0)$.
The spatial dependence of the $\Delta$ in an SNS junction is shown
schematically in Fig. 4. The value of the $\Delta$ on both side of
the interface are related through the standard boundary condition
\cite{Parks,Tinkham}:

\begin{figure}
\includegraphics[height=7cm,width=9cm]{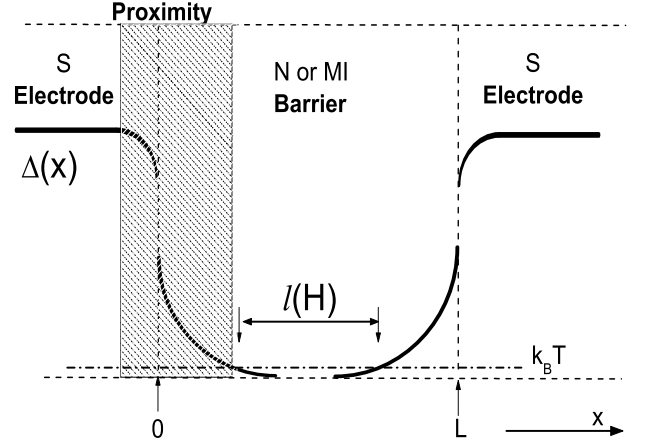}
\caption{\label{fig:epsart} A schematic diagram of the junction
and the spatial profile of $\Delta(x)$. $\ell(H)$ is the length of
the resistive region of the junction. The shaded area shows the
region in which superconductivity is weakened on both sides of the
interface due to the proximity effect.}
\end{figure}

\begin{equation}\label{4}
\left(\frac{\Delta_S^i}{N_{S}V_{S}}\right)_{x=0}=\left(\frac{\Delta_N^i}{N_{N}V_{N}}\right)_{x=0}
\end{equation}

\noindent where $\Delta_S^i$ and $\Delta_N^i$ are the values of
the minigap at the S and N sides of the SN interface.  $N_{S}$ and
$N_{N}$ are the normal state density of states (DOS) on the S and
N sides of the interface, respectively. Finally, $V_{S}$ and
$V_{N}$ are the electron-electron interaction on the S and N
sides. Assuming the DOS and the electron-electron interaction are
field independent we obtain

\begin{equation}\label{5}
\frac{\Delta_S^i(H)}{\Delta_N^i(H)}=\epsilon=\frac{\delta_S^i}{\delta_N^i}
\end{equation}

\noindent where $\epsilon=N_{S}V_{S}/N_{N}V_{N}$ is a field
independent constant and we define
$\delta_N\equiv\Delta_N(0)-\Delta_N(H)$. $\delta_N^i$ which is the
value at the interface, is also a small
parameter as $\delta_N^i/\Delta_N^i(0)<<1$.\\

Turning to the N side now, under a magnetic field H applied in the
c-direction, the spatial dependence of $\Delta$ is given by the
linearized Ginzburg-Landau (GL) equation \cite{Parks}

\begin{equation}\label{6}
-\frac{d^2\Delta_N}{dx^2}+\left(\frac{2eH}{\hbar
c}\right)^2(x_0-x)^2\Delta_N+K^2\Delta_N=0
\end{equation}

\noindent where $x_0$ and K are constants.  In our experiment,
$x_0-x$ is limited by 10 nm which is half the thickness of our
junction, the field H is less than 8T, and K is on the order of a
few nm. Using these parameters, we estimate that the upper limit
of the second term in Eq. (8) is about two orders of magnitude
smaller than the last term. In this limit, the solution of Eq. (8)
for $\Delta$ in N exhibits an exponential decay with distance
$\Delta_N(x)=\Delta_N^iexp(-Kx)$. In the dirty limit \cite{Parks},
K is given by

\begin{equation}\label{8}
K^{-1}=\left(\frac{\hbar D_{N}}{2\pi k_{B}T}\right)^{\frac{1}{2}}.
\end{equation}

\noindent In this limit, where the cyclotron radius in the
magnetic field is much larger than the mean free path, $D_{N}$ is
field independent and thus K does not depend on field. However,
the value of $\Delta$ at the interface $\Delta_N^i$, is field
dependent because it is pinned to the value of the $\Delta$ on the
S side at the interface $\Delta_S^i$ through Eq. (5). The pair
amplitude induced in the barrier is effectively depressed to zero
by thermal fluctuations at some distance from the interface, and
from that distance onwards the material has a finite resistance.
The natural way to determine this distance is through the
condition that the extrapolated magnitude of $\Delta$ there is of
the order of $k_{B}T$. This length, which we denote by $X$,
depends on the field as

\begin{equation}\label{9}
k_{B}T=\Delta_N^i(H)e^{-KX(H)}
\end{equation}

\noindent where $X(H)$ is the effective penetration depth of
superconductivity into N when a magnetic field is applied.
Dividing the Eq. (9) for $X(H)$ by the equation for $X(H=0)$ we
find

\begin{equation}\label{10}
\frac{\Delta_N^i(0)}{\Delta_N^i(H)}=e^{K[X(0)-X(H)]}
\end{equation}

\noindent and

\begin{equation}\label{11}
X(H)-X(0)=\frac{1}{K}ln\left(1-\frac{\delta_N^i}{\Delta_N^i(0)}\right).
\end{equation}

\noindent Since $\delta_N^i$ is a small parameter $X(H)-X(0)\sim
-\delta_N^i/K\Delta_N^i(0)$. Referring to the schematic model
shown in Fig. 4, the field dependent resistance of the barrier is
$R=\rho \ell(H)/A$, where $\ell(H)=L-2X(H)$ is the length inside
the barrier which is normal. Using Eq. (5) and the relation
$2\Delta_S=\beta k_B T_c$, the magnetoresistance comes out as

\begin{eqnarray}\label{16}
MR&\equiv&R(H)-R(0)=-2\frac{\rho}{A}(X(H)-X(0))\\
&=&\frac{\pi \rho eD_{S}}{2cAK}\left(\frac{1}{k_BT_c}\right) H
\nonumber
\end{eqnarray}

\noindent We therefore see that the MR is linear in H, in
agreement with the observed behavior in Fig. 2.\\

A rough estimate of the decay length ($1/K$) in the underdoped
barrier at low temperature can be attempted using the resistivity
of the barrier, $0.8\Omega cm$ and the typical resistance of the
junctions $\sim 100\Omega$. Using these values, we estimate the
length of the barrier which remains normal $\ell(H=0)$ as 6nm.
Taking the thickness of the barrier of 21nm, and assuming that the
pair amplitude decays to zero over 3 times the decay length
($1/K$), we obtain a value for $1/K$ of about 2.5nm. $1/K$ can
also be calculated using Eq. (7), where
$D_N=\frac{1}{3}\ell_{N}v_{FN}$. The mean free path in the barrier
can be estimated as the distance between nearest Ga atoms
$\ell_{N}\sim 5\rm \AA$ and the Fermi velocity in the barrier
$v_{FN}=1.2\times 10^7cm/sec$ was measured in a previous
study\cite{KorenMRO}. This yields $1/K\simeq $3.5nm. It appears
that both methods of estimating $1/K$ give values which are
consistent. We note that the decay length estimated in underdoped
SNS structure comes out much smaller than in
optimally doped ones \cite{Kleinsasser,Polturak}.\\

Using our data we estimate the diffusion coefficient $D_{S}$ and
the relaxation time $\tau_{S}$ of underdoped YBCO. Taking an
average $T_c$ of $\sim$45K, and an average slope in Fig. 3 of
$MR/H=2.7\Omega/Tesla$,  Eq. (11) yields $D_{S} \sim 1
cm^{2}/sec$. This is consistent with an independent estimate that
can be extracted from Eq. (2) and from the relation between
$\alpha$ and the diffusion coefficient $D_S$ which yields $\sim
1.7 cm^{2}/sec$. The relaxation time, $\tau_{S}$ is extracted from
the usual relation that connects it with the diffusion coefficient
$D_{S}=\frac{1}{3} v_{FS}^{2} \tau_{S}$ where $v_{FS}$ is the
Fermi velocity of quasiparticles in the superconductor $\sim
2\times 10^{7} cm/sec$ \cite{Orenstein}. Under these assumptions
$\tau_{S}$ for YBCO is $\sim 25fs$. The value found for $\tau_{S}$
is of the same order of magnitude as the recent results of Gedik
\textit{et al.} who obtained $\tau_{S}\sim 100\,fs$
\cite{Orenstein} while our value of $D_S$ is smaller by than
theirs, $D_S\sim 20$ cm$^2$/s.\\

For completeness, we mention that Abrikosov has predicted another
mechanism for linear MR versus H in superconductors
\cite{Abrikosov}. He assumed a field dependent resonant tunneling
which yields MR linear in H at very high magnetic fields, when
only a few Landau levels are filled. When the field is reduced and
the number of filled Landau levels increases the field dependence
of the MR changes into a quadratic one. This model could  in
principle, explain the observed linear behavior of our MR results.
However, peaks in the density of states due to Landau levels are
absent in the dynamic resistance spectra of our junctions.
Moreover, the fields used in our experiment are not high enough to
reach the regime where a low number of Landau levels are filled.
Hence, if this model was applicable to our junctions, we should
have observed a quadratic dependence of the MR on field, which
is not the case.\\

 \maketitle{\large CONCLUSIONS}

We investigated the resistance of SNS structures based on
underdoped YBCO with non-magnetic Ga-doped YBCO barrier as a
function magnetic field. We discovered a linear increase of the
resistance with the field MR. An extensive series of control
experiments indicates that this field dependence does not result
from flux flow, which would be the obvious mechanism of MR in a
superconductor. A simplified analysis indicates that the effect
may well be explained by field dependent proximity effect in the
barrier. This explanation produces a reasonable estimate of the
diffusion coefficient and the relaxation time in YBCO.
Furthermore, our estimates indicate that in underdoped YBCO SNS
structures, the superconductivity induced inside the barrier
through the proximity effect has a short ($\sim$2-3 nm) range,
unlike the long range proximity effect observed in
optimally doped YBCO structures.\\

{\em Acknowledgments:} We thank Pavel Aronov for the "short"
junction data of Fig. 2. This research was supported in part by
the Israel Science Foundation (grant No. 1565/04), the Heinrich
Hertz Minerva Center for HTSC, the Karl Stoll Chair in advanced
materials, and by the Fund for the Promotion of Research at the
Technion.\\

\subsection{\large APPENDIX}

\begin{figure}
\includegraphics[height=7cm,width=9cm]{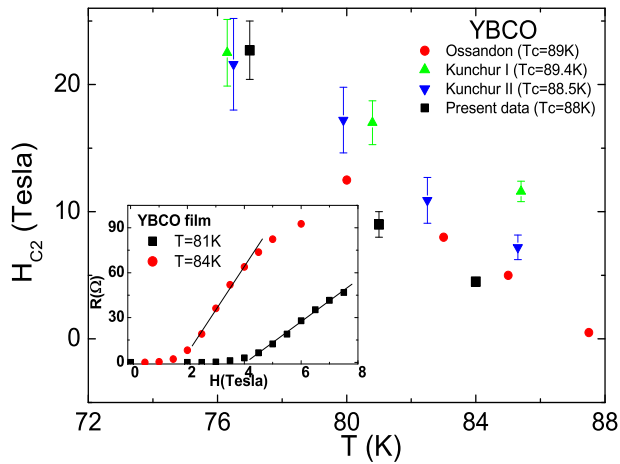}
\caption{\label{fig:epsart} (Color online) $H_{c2}$ versus
temperature of optimally doped YBCO film. $H_{c2}$ was extracted
from the slope of the linear part of the MR (inset) using the
Bardeen-Stephen model. Our data (solid squares) can be compared
with previous measurements by Kunchur \textit{et al.} and Ossandon
$\tau_{S}$ \cite{phillips,Ossandon}.}
\end{figure}

For the sake of comparison with previous work, we also measured
the MR of \textit{optimally doped} YBCO films at temperatures
close to $T_c$ as shown in the inset of Fig. 5. In this case, the
resistance showed a region linear with applied field. In the
Bardeen-Stephen model \cite{BardeenStephan,Tinkham,Kim}, the
resistance resulting from flux flow is given by $R_{flux flow
}=(H/H_{c2}(T))\times R_{N}(T)$ where H is the applied magnetic
field and $R_{N}(T)$ is the normal state resistance at temperature
T, extrapolated from the RT plot close to $T_c$. Using this model,
we extracted the temperature dependence of $H_{c2}$ near $T_{c}$.
Our results show good agreement with previous measurements by
Kunchur \textit{et al.} and Ossandon {\em et al.}
\cite{phillips,Ossandon}, which are also plotted in  Fig. 5. At
temperatures much lower than $T_c$ however, no measurable MR in
the YBCO film was observed. Therefore, at low temperatures where
the junctions of Fig. 3 were measured, flux flow in the YBCO
electrodes does not contribute to the MR. This conclusion holds
independent of the oxygen doping level of the YBCO.\\



\end{document}